\documentclass[pra,twocolumn,superscriptaddress,amsmath,amssymb,floatfix]{revtex4-1}
\usepackage{graphicx}
\usepackage{eqnarray}

\begin{document}

Published in: Phys. Rev. B \textbf{103}, 115206 (2021).\\

\begin{flushright}
DOI: 10.1103/PhysRevB.103.115206\\
\end{flushright}

\title{Broadband optical conductivity of the chiral multifold semimetal PdGa}

\author{L. Z. Maulana}
\affiliation{1.~Physikalisches Institut, Universit\"at Stuttgart,
70569 Stuttgart, Germany}
\author{Z. Li}
\affiliation{MIIT Key Laboratory of Advanced Display Materials and
Devices, Ministry of Industry and Information Technology, Institute
of Optoelectronics and Nanomaterials, Nanjing University of Science
and Technology, Nanjing 210094, China}
\author{E. Uykur}
\affiliation{1.~Physikalisches Institut, Universit\"at Stuttgart,
70569 Stuttgart, Germany}
\author{K. Manna}
\affiliation{Max-Planck-Institut f\"ur Chemische Physik fester
Stoffe, 01187 Dresden, Germany} \affiliation{Department of Physics,
Indian Institute of Technology Delhi, Hauz Khas, New Delhi 110016,
India}
\author{S. Polatkan}
\affiliation{1.~Physikalisches Institut, Universit\"at Stuttgart,
70569 Stuttgart, Germany}
\author{C. Felser}
\affiliation{Max-Planck-Institut f\"ur Chemische Physik fester
Stoffe, 01187 Dresden, Germany}
\author{M. Dressel}
\affiliation{1.~Physikalisches Institut, Universit\"at Stuttgart,
70569 Stuttgart, Germany}
\author{A. V. Pronin}
\email{artem.pronin@pi1.physik.uni-stuttgart.de}
\affiliation{1.~Physikalisches Institut, Universit\"at Stuttgart,
70569 Stuttgart, Germany}

\date{March 29, 2021}

\begin{abstract}

We present an optical conductivity study of the multifold semimetal
PdGa, performed in a broad spectral range (100 -- 20\,000~cm$^{-1}$;
12 meV -- 2.5 eV) down to $T = 10$~K. The conductivity at
frequencies below 4\,000~cm$^{-1}$ is dominated by free carriers
while at higher frequencies interband transitions provide the major
contribution. The spectra do not demonstrate a significant
temperature evolution: only the intraband part changes as a function
of temperature with the plasma frequency remaining constant. The
interband contribution to the conductivity exhibits a broad peak at
around 5\,500~cm$^{-1}$ and increases basically monotonously at
frequencies above 9\,000~cm$^{-1}$. The band-structure-based
computations reproduce these features of the interband conductivity
and predict its linear-in-frequency behavior as frequency
diminishes.

\end{abstract}

\maketitle

\section{Introduction}\label{Introduction}

The solid-state realization of massless Weyl fermions was first
theoretically discussed~\cite{Murakami2007, Wan2011} and
subsequently experimentally confirmed in TaAs~\cite{Lv2015, Xu2015},
one of the materials belonging to a broad, by now, class of solids
known as topological Weyl semimetals. Since these discoveries,
various families of other topological semimetals -- materials with
the nontrivial band topology relevant for the bulk states -- were
unveiled~\cite{Burkov2011, Zhu2016, Bzdusek2016, Burkov2016,
Armitage2018}. One particular example is multifold semimetals. In
these compounds, the topologically protected band crossings of
degeneracy higher than two are described by Weyl-like Hamiltonians:
linear in momentum and in effective spin, which can be larger than
1/2 \cite{Manes2012, Bradlyn2016}.

Recently, a number of compounds from the cubic space group 198
(SG198), which is non-centrosymmetric and has no mirror planes, were
confirmed to possess such multifold crossing points~\cite{Chang2018,
Chang2017, Tang2017, Sanchez2019, Rao2019, SchroeterAlPt,
SchroeterPdGa, Takane2019, Yao2020}. In these materials, the band
crossings with different chiralities are situated at different
energy positions, providing thus a realization of electronic chiral
crystals. Remarkably, the crystals of these compounds can be grown
as single enantiomers, i.e., with a given crystalline and electronic
chirality \cite{SchroeterPdGa, Sessi2020}.

Generally, optics seems to be an ideal tool for studying chiral
materials and for manipulating the chiral degrees of freedom, as the
circularly polarized light can be directly coupled to the chiral
quasiparticles within such solids. For such investigations, it is
essential to possess an advance knowledge of the linear optical
response for the materials of interest. In other families of
topological semimetals, optical spectroscopy, a genuine
bulk-sensitive technique, was shown to provide insight into the bulk
electronic properties~\cite{Armitage2018, Pronin2020}. Thus, efforts
were recently taken to calculate and to measure the
frequency-dependent conductivity,
$\sigma(\omega)=\sigma_{1}(\omega)+i\sigma_{2}(\omega)$, of the
chiral multifold compounds. Specifically, the optical conductivity
of two members of this family, RhSi and CoSi, was studied
theoretically~\cite{Li2019, Habe2019} and
experimentally~\cite{Marel1998, Rees2019, Maulana2020, Xu2020,
Ni2020}. Linear-in-frequency conductivity $\sigma_{1}(\omega)$ due
to interband transitions at low frequencies was evidenced in both
RhSi and CoSi, consistent with theory predictions for multifold
semimetals~\cite{Grushin2019} and with general expectations for
interband optical conductivity in linearly dispersing
three-dimensional electronic bands~\cite{Hosur2012, Bacsi2013,
Ashby2014}. In this paper, we expand the experimental studies of
$\sigma(\omega)$ to another multifold semimetal, PdGa. Recently,
this compound was thoroughly studied by ARPES~\cite{SchroeterPdGa}
and scanning tunneling microscopy \cite{Sessi2020}, and a very high
Chern number, $\mid C \mid = 4$, was evidenced in these studies.

\section{Experiment}\label{Experiment}

PdGa single crystals were grown from its melt by a self-flux
technique, as described in Refs.~\cite{SchroeterPdGa, Sessi2020}.
First, a polycrystalline ingot was prepared using the arc melt
technique with a stoichiometric mixture of high-purity Pd and Ga.
Then, the crushed powder was filled in a thin-wall alumina crucible
and finally sealed in a quartz ampoule. The crystal growth was done
under a partial vacuum of 3 mbar. The ampoules were heated to
1100~$^{\circ}$C, kept there for 12 h, and then slowly cooled to
900~$^{\circ}$C at a rate of 1.5~$^{\circ}$C/h. Finally, the samples
were cooled to 800~$^{\circ}$C at a rate of 50~$^{\circ}$C/h,
annealed for 120 h and then cooled to 500~$^{\circ}$C at a rate of
5~$^{\circ}$C/h. PdGa single crystals with average linear dimensions
of a few mm were obtained.

The crystals were first analyzed with a white beam backscattering
Laue x-ray diffractometer at room temperature. The obtained single
and sharp Laue spot could be indexed by a single pattern, revealing
the excellent quality of the grown single-enantiomer crystals
without any twinning or domains, see the Supplemental Material below
for a Laue-pattern example. The structural parameters were
determined using a Rigaku AFC7 four-circle diffractometer with a
Saturn 724+ CCD-detector applying graphite-monochromatized
Mo-K$\alpha$ radiation. The crystal structure was refined to be
cubic $P2_{1}3$ (SG198) with a lattice constant $a$=4.896~{\AA}.

\begin{figure}[t]
\centering
\includegraphics[width=\columnwidth,clip]{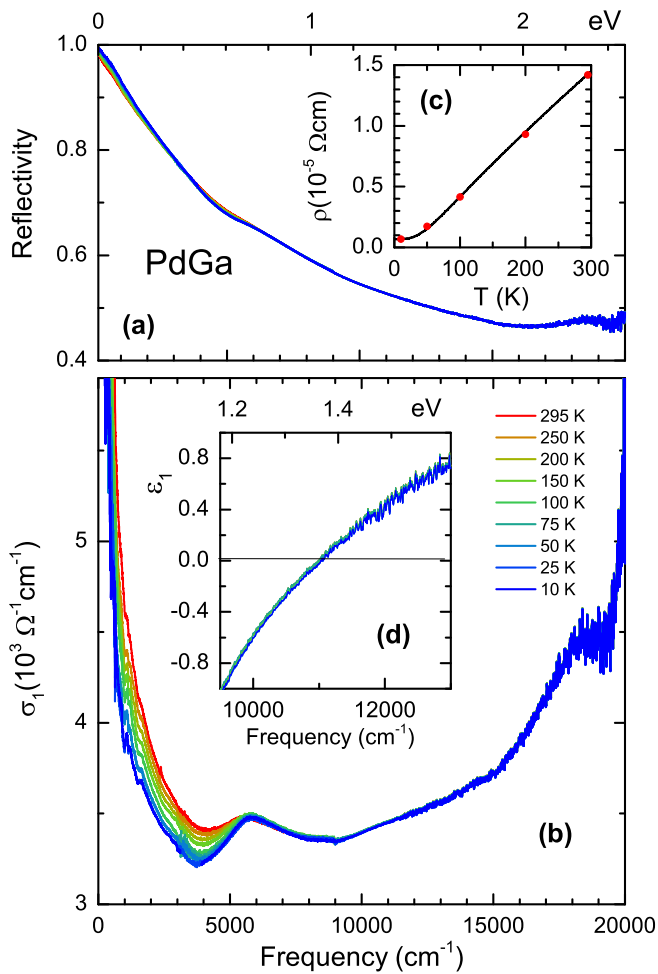}
\caption{PdGa optical reflectivity (a) and the real part of optical
conductivity (b) at selected temperatures as indicated. The insets
show (c) the dc resistivity vs $T$ (line) together with the inverse
optical conductivity in the $\nu \rightarrow 0$ limit used in the
fits of Fig.~\ref{fit} (bold dots) and (d) zoomed permittivity
spectra near the zero-line crossing. Note that in (a), (b), and (d),
the experimental curves are presented for all indicated
temperatures.} \label{overview}
\end{figure}

Temperature-dependent transport measurements (longitudinal dc
resistivity and Hall) were performed in custom-made setups at
temperatures down to 2 K.

Temperature-dependent ($T$ = 10 -- 295 K) optical reflectivity,
$R(\nu)$, was measured on a PdGa single crystal (with roughly $1.5
\times 1.5$ mm$^{2}$ in lateral dimensions) over a broad frequency
range from $\nu = \omega/(2 \pi)= 100$ to 20\,000 cm$^{-1}$ (12 meV
-- 3 eV). The spectra in the far-infrared (below 700 cm$^{-1}$) were
collected with a Bruker IFS 113v Fourier-transform spectrometer
using an \textit{in situ} freshly evaporated gold over-filming
technique for reference measurements. At higher frequencies, a
Bruker Hyperion infrared microscope attached to a Bruker Vertex 80v
FTIR spectrometer was used. For these measurements, freshly
evaporated gold mirrors on glass substrates served as the reference.
No sample anisotropy was detected, which is in agreement with the
cubic crystallographic structure.

For the Kramers-Kronig analysis, the zero-frequency extrapolations
were made using the Hagen-Rubens relation in accordance with
temperature-dependent longitudinal dc resistivity measurements. For
high-frequency extrapolations, we utilized x-ray atomic scattering
functions~\cite{Tanner2015} followed by the free-electron behavior,
$R(\omega) \propto 1/\omega^{4}$, above 30 keV.

\begin{figure}[t]
\centering
\includegraphics[width=\columnwidth,clip]{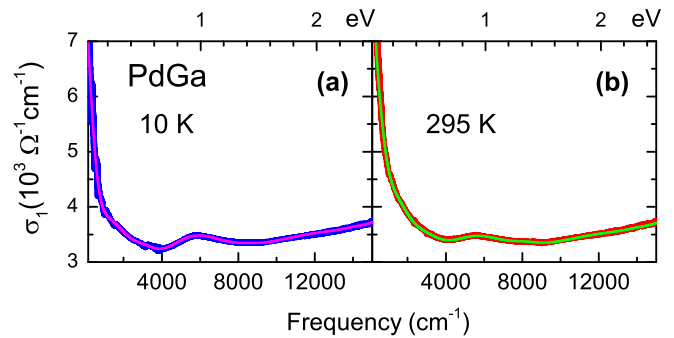}
\caption{Drude-Lorentz fits (lines) of the measured optical
conductivity spectra (symbols) at 10 and 295 K.} \label{fit}
\end{figure}

\section{Calculations}\label{Calculations}

The band structure and optical properties of PdGa were calculated by
first-principles calculations based on the density-functional theory
with the Perdew-Burke-Ernzerhof exchange-correlation functional
implemented in Quantum ESPRESSO~\cite{QE}. Norm-conserving
pseudopotentials with the generalized gradient approximation (GGA)
are adopted in this work. We used the experimental structural
parameters~\cite{Sessi2020} without any geometry optimization. The
energy cutoff of $35.0 R_{y}$ and a $k$-point grid with
24$\times$24$\times$24 (40$\times$40$\times$40) points were adopted
for the band-structure (optical-properties) calculations. The
spin-orbit-coupling effect as well as a frequency-independent
broadening for interband transitions due to electron scattering (0.2
eV) were included in the calculations.

\section{Results and Discussion}\label{Results}

Examples of the frequency-dependent optical spectra are shown in
Fig.~\ref{overview} for selected temperatures. The raw reflectivity
shows typical metallic behavior with $R(\nu)$ approaching unity as
frequency diminishes, see panel (a). The spectra of the real part of
optical conductivity [panel (b)] demonstrate corresponding behavior:
$\sigma_{1}$ increases as $\nu \rightarrow 0$. This is in
qualitative agreement with a simple free-electron Drude model. The
screened plasma frequency of free electrons, $\nu_{pl}^{scr}$, can
be estimated from the zero crossings of the permittivity spectra,
$\varepsilon_{1}(\nu)=1-2\sigma_{2}(\nu)/\nu$, shown in panel (d).
We found $\nu_{pl}^{scr}$ to be temperature independent at
$\nu_{pl}^{scr} \simeq 11\,000$~cm$^{-1}$, corresponding to
$\hslash\omega_{pl}^{scr}=1.37$ eV.

\begin{figure}[b]
\centering
\includegraphics[width=\columnwidth,clip]{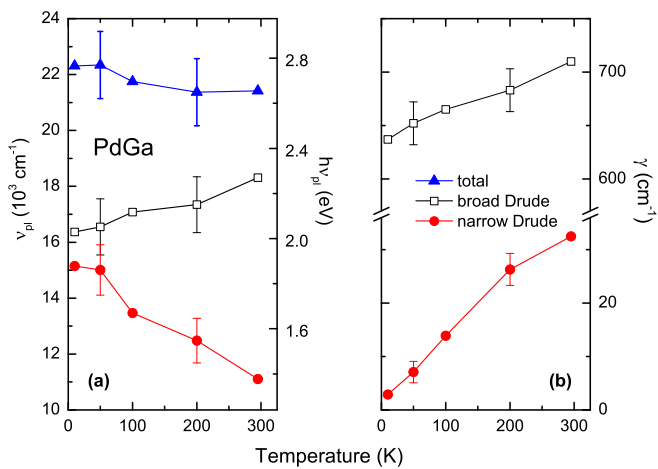}
\caption{Temperature-dependent parameters of the two Drude terms
(narrow and broad) used in the Drude-Lorentz fits. Note different
vertical scales for the plasma frequencies (a) and the scattering
rates (b).} \label{Drude}
\end{figure}

Comparing these findings with the experimental results on
$\nu_{pl}^{scr}$ in many other nodal semimetals~\cite{Marel1998,
Maulana2020, Xu2020, Chen2015, Xu2016, Neubauer2016, Frenzel2017,
Kimura2017, Huett2018}, one can immediately notice a very large
value of $\nu_{pl}^{scr}$ in PdGa. For example, it is roughly one
order of magnitude larger, than in the sister compounds, the
multifold semimetals RhSi, $\nu_{pl}^{scr} \sim 1500$ --
$1700$~cm$^{-1}$ \cite{Maulana2020, Ni2020}, and CoSi,
$\nu_{pl}^{scr} \sim 600$ -- $800$~cm$^{-1}$ \cite{Marel1998,
Xu2020}. This ``metallicity'' of PdGa is obviously related to its
band structure with the chemical potential situated far from the
nodes, as discussed below. Our Hall measurements (see the
Supplemental Material) also reveal the metallic nature of PdGa: its
electron density is high and almost temperature-independent, $n = (3
\pm 1) \times 10^{22}$ cm$^{-3}$.

As seen in Fig.~\ref{overview}, the overall temperature evolution of
the conductivity spectra is rather weak: only the low-energy
free-electron part shows detectable $T$-induced changes due to the
temperature-dependent electron scattering, in agreement with the
metal-like optical response. At $\nu > 4000$~cm$^{-1}$, the
interband transitions start to become visible in $\sigma_{1}(\nu)$.

To analyze the optical spectra in a more quantitative way, we first
performed a standard Drude-Lorentz fit~\cite{Dressel2002} for a
number of temperatures. The Drude contribution describes the
intraband response, while the Lorentzians are used to fit the
interband optical transitions. Examples of such fits are presented
in Fig.~\ref{fit}. In all the fits, we kept the zero-frequency limit
of optical conductivity equal to the measured dc-conductivity value
at every temperature, see Fig.~\ref{overview}(c). No other
restrictions on the fit parameters were imposed. The fits were
obtained by the simultaneous fitting of $R(\nu)$, $\sigma_{1}(\nu)$
and $\varepsilon_{1}(\nu)$.

\begin{figure}[t]
\centering
\includegraphics[width=0.9\columnwidth,clip]{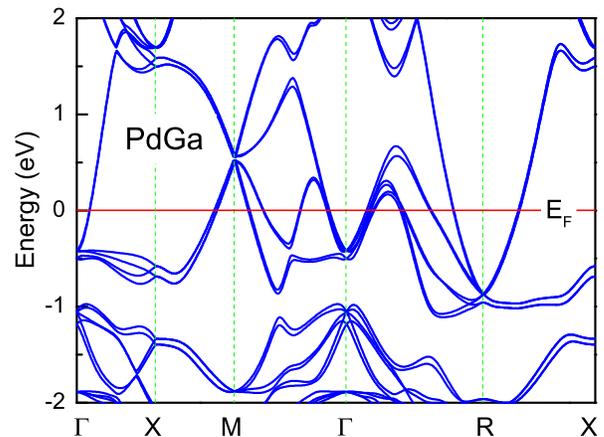}
\caption{Low-energy electronic band structure of PdGa with the
spin-orbit coupling included. The multifold fermions are supposed to
exist near the $\Gamma$ and $R$ points.} \label{bands}
\end{figure}

\begin{figure}[b]
\centering
\includegraphics[width=0.9\columnwidth,clip]{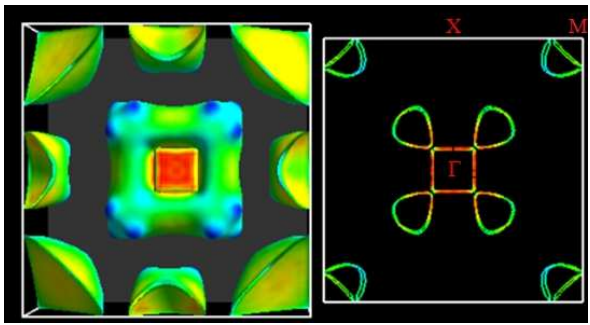}
\caption{Calculated bulk Fermi surface of PdGa (left) and its cut at
the middle of the Brillouin zone (right). The $k_{z}$ direction is
perpendicular to the picture plane.} \label{FS_PdGa}
\end{figure}

We found that we need at least two Drude components (``narrow'' and
``broad'') with different scattering rates to provide accurate
Drude-Lorentz fits to the experimental spectra. This approach is
often used to describe the intraband optical response in different
multiband materials~\cite{Maulana2020, Ni2020, Wu2010,
Schilling2017, Neubauer2018, Kemmler2018, Qiu2019, Yang2020,
Homes2020}, but the exact interpretation of the two Drude terms
remains arguable~\cite{Maulana2020, Ni2020, Kemmler2018}. PdGa
possesses not two, but many different bands crossing the Fermi
level, see Figs.~\ref{bands} and \ref{FS_PdGa}. Hence, the two
components might be associated, e.g., with two different
\textit{sets} of bands, or with different scattering mechanisms. In
any case, the two-Drude approach utilized here should be considered
just as a minimalist model to fit the intraband optical response in
a Kramers-Kronig consistent way and to extract the total spectral
weight (the plasma frequency) of itinerant carriers.

The fit parameters of the Drude terms are shown in Fig.~\ref{Drude}
as functions of temperature. Because of the temperature-induced
redistribution of the spectral weight between the terms, neither of
the scattering rates ($\gamma$) is expected to follow the
dc-resistivity temperature dependence accurately. Nevertheless, it
is primarily the temperature variation of the narrow-Drude
scattering rate, which provides the reconciliation of the relatively
strong temperature dependence of dc conductivity [the residual
resistivity ratio is around 20, see Fig.~\ref{overview}(c)] and the
fairly weak temperature evolution of intraband optical conductivity;
see also the Supplemental Material.

While the parameters of the Drude terms vary appreciably with
temperature, the total intraband unscreened plasma frequency
$\nu_{pl}$ remains almost temperature independent at $\sim 21\,800$
cm$^{-1}$ ($\hbar\omega_{pl} \approx 2.7$ eV). This value is
significantly larger than the one reported for RhSi ($\sim
11\,300$~cm$^{-1}$ or 1.4~eV) \cite{Maulana2020}, that is in
qualitative agrement with the smaller bulk Fermi surface of RhSi,
see the Supplemental Material. Also, the fit-based $\nu_{pl}$ in
PdGa is consistent with the screened plasma frequency
$\nu_{pl}^{scr}$, obtained from zero crossings of
$\varepsilon_{1}(\nu)$, if the higher-frequency dielectric constant
$\varepsilon_\infty$ is assumed to be around 1.4, which is a
reasonable value [cf. Fig.~\ref{overview}~(d)]. We recall that
$\nu_{pl}^{scr} = \nu_{pl} / \sqrt{\varepsilon_\infty}$.

\begin{figure}[t]
\centering
\includegraphics[width=\columnwidth,clip]{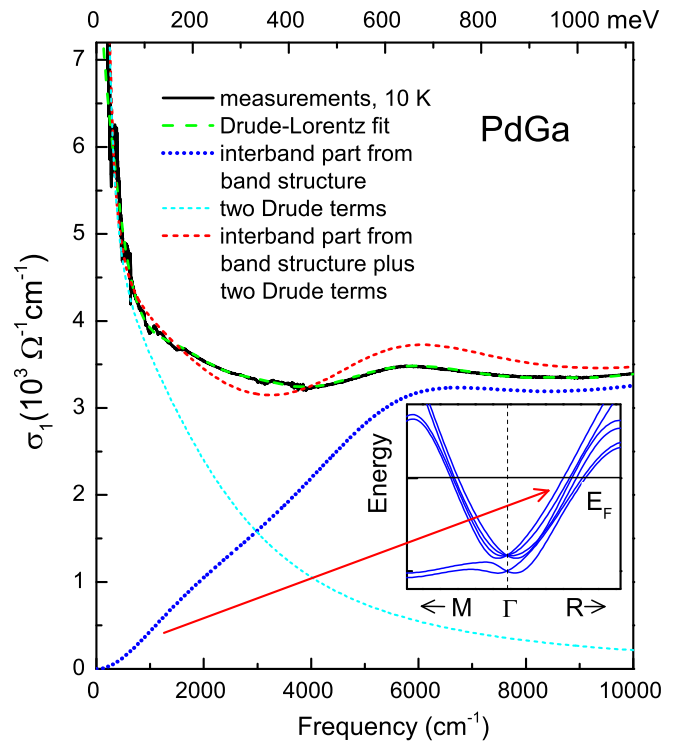}
\caption{Comparison of the measured (solid line) and calculated
(dashed and dotted lines) optical conductivity in PdGa. The
interband portion of $\sigma_{1}(\omega)$ (blue dotted line) is
calculated from the band structure. Adding two Drude terms (cyan
dashed line) to this curve provides a good qualitative match (red
dashed line) to the experimental spectrum. The green dashed line is
the fit from Fig.~\ref{fit}. The quasilinear interband conductivity
at low frequencies is due to the transitions between the multiple
bands in the vicinity of the Fermi level, as shown in the inset.}
\label{comparison}
\end{figure}

For a more elaborative analysis, we performed band-structure
calculations for PdGa and then computed its interband optical
conductivity, as described above. The results of the band-structure
calculations are shown in Fig.~\ref{bands}. It is apparent that the
Fermi level $E_{F}$ crosses a number of electron- and hole-like
bands and that the multifold nodes near the $\Gamma$ and $R$ points
are situated quite deep (0.5 eV or more) below $E_{F}$. Hence, PdGa
possesses an extended bulk Fermi surface, which occupies a
significant portion of the Brillouin zone, see Fig.~\ref{FS_PdGa}.

The large bulk Fermi surface and the consequent intense intraband
optical response make a direct comparison between the interband
conductivity computed from the band structure and the experimental
spectra challenging. As seen from Fig.~\ref{comparison}, the
interband portion of optical conductivity is not zero down to very
low frequency. This might look surprising, as usually there is an
onset (a Pauli edge) in the interband-conductivity spectra for
systems with the Fermi level situated far from the nodes, which is
the case for PdGa. However, PdGa possess multiple bands with small
energy separation, which (the bands) cross the Fermi level, see
Fig.~\ref{bands} and the inset of Fig.~\ref{comparison}. The
cumulative effect of the transitions between these bands causes the
quasilinearity of the interband $\sigma_{1}(\nu)$ in PdGa at low
frequencies. We note that at $\nu \rightarrow 0$, $\sigma_{1}(\nu)$
flattens out. More details on band-selective optical transitions in
PdGa can be found in Ref.~\cite{Polatkan2021}.

In order to compare the experimental and computed interband
conductivity, the intraband (Drude) response can be subtracted from
the former spectra, as it was done, e.g., in
Refs.~\cite{Maulana2020, Ni2020, Chaudhuri2017, Corasaniti2019}.
This procedure may obviously produce ambiguities in determining the
interband portion of the experimental $\sigma_{1}(\nu)$, especially
at the lowest frequencies, where the interband conductivity is low.
Hence, we followed a slightly different approach: instead of
subtracting the Drude terms from the experimental spectra, we added
them to the calculated interband conductivity. Basically, we
performed a sort of fit with two Drude terms and the interband
$\sigma_{1}(\nu)$ obtained from the band structure with a
frequency-independent electron scattering rate of 0.2 eV ($\sim$
1600 cm$^{-1}$). The results of this analysis are shown in
Fig.~\ref{comparison} as well as in the Supplemental Material. As a
starting point, we used the Drude terms, obtained from our
Drude-Lorentz fits (Fig.~\ref{fit}). For the best possible
description of the experimental spectra, we had to slightly change
the parameters of these terms, but the zero-frequency limit of
$\sigma_{1}(\nu)$ remained to be equal to the inverse of the
measured dc resistivity.

Having in mind that the band-structure-based computations of
$\sigma_{1}(\nu)$ generally reproduce the experimental findings only
qualitatively~\cite{Pronin2020, Maulana2020, Neubauer2016,
Kimura2017, Huett2018, Chaudhuri2017}, the match between theory and
experiment can be considered as rather good: in the present case the
general conductivity level observed in experiment is reproduced by
computations and the major feature of the interband
$\sigma_{1}(\nu)$ -- the flat maximum at around 5500~cm$^{-1}$ -- is
seen in both computed and measured spectra. The remaining
discrepancies can be attributed, e.g., to a frequency-dependent
electron scattering in the investigated sample (as noticed above, we
assumed a frequency-independent scattering rate in our
computations).

Due to the very high free-carrier concentration and relatively large
electron scattering (the broad Drude component), direct experimental
verification of the linear interband optical conductivity at low
energies, as it is predicted by theory for multifold semimetals with
the nodes situated in the vicinity of $E_{F}$ \cite{Grushin2019}, is
impossible for PdGa. Nevertheless, our experimental spectra can be
well described as a sum of the interband conductivity, obtained from
the band structure, and a strong Drude-like free-carrier
contribution. Let us also note that this strong electronic response
prevents observation of any phonon modes on the top of it (based on
its crystallographic symmetry, PdGa is supposed to have five
infrared-active phonons).

\section{Conclusions}\label{Conclusions}

We have studied the broadband optical conductivity of the multifold
semimetal PdGa. A prominent metallic response is detected. The
free-carrier Drude-like contribution with a temperature-independent
plasma frequency ($\hslash\omega_{pl}^{scr}=1.37$ eV) dominates the
spectra at frequencies below 4000 cm$^{-1}$, preventing direct
detection of the linear-in-frequency interband conductivity
predicted for the multifold semimetals. At higher frequencies, the
spectra calculated from the band structure reproduce the
experimental spectra. Namely, the general conductivity levels
obtained in experiments and in computations match each other and the
frequency position of the most prominent feature in the experimental
interband conductivity -- the maximum at around 5500~cm$^{-1}$ (680
meV) -- is reproduced by the computations.

\section{Acknowledgments}

We thank Gabriele Untereiner for valuable technical support. E.U.
acknowledges financial support from the European Social Fund and
from the Ministry of Science, Research, and the Arts of
Baden-W{\"u}rttemberg. K.M. and C.F. acknowledge financial support
from the European Research Council (ERC) via Advanced Grant No.
742068 ``TOP-MAT''. The work in Stuttgart was partly supported by
the Deutsche Forschungsgemeinschaft (DFG) via Grant No. DR228/51-3.

\section{Supplemental Material}

\begin{figure}[h!]
\centering
\includegraphics[width=7 cm,clip]{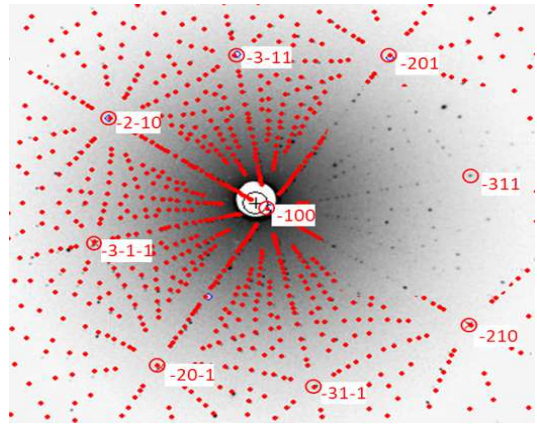}
\caption{Laue pattern of a PdGa single crystal, superposed with a
simulated pattern.} \label{Laue}
\end{figure}

\begin{figure}[h!]
\centering
\includegraphics[width=5.5 cm,clip]{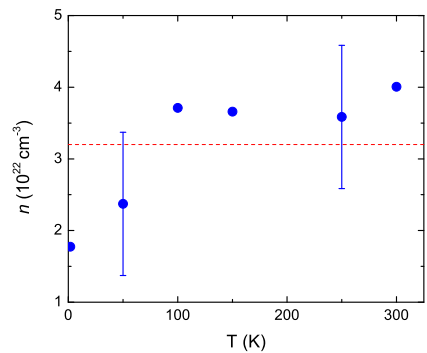}
\caption{Carrier concentration in the PdGa sample used in the
optical experiments versus temperature, as obtained form Hall
measurements. The dashed horizontal line shows the averaged value.}
\label{Hall}
\end{figure}

\textbf{Material and sample characterization.} The results of our
Laue and Hall measurements are presented in Figs.~\ref{Laue} and
\ref{Hall}.

\begin{figure}[]
\centering
\includegraphics[width=7 cm,clip]{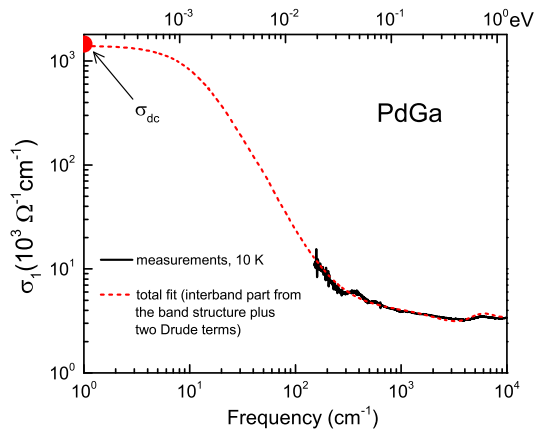}
\caption{Same graphs as shown in Fig.~\ref{comparison} (the 10-K
conductivity data and its fit) on a double-logarithmic scale. The dc
conductivity value is shown as a bold red dot on the left vertical
axis.} \label{log_fits}
\end{figure}

\textbf{Two-component Drude fits.} In order to demonstrate how the
low-frequency portion of the measured optical conductivity and the
dc-conductivity value are reconciled in our fits, we re-plot the
10-K experimental spectrum and its fit shown in
Fig.~\ref{comparison} on a double-logarithmic scale in
Fig.~\ref{log_fits}.\\

\begin{figure}[]
\vspace{1cm} \centering
\includegraphics[width=\columnwidth,clip]{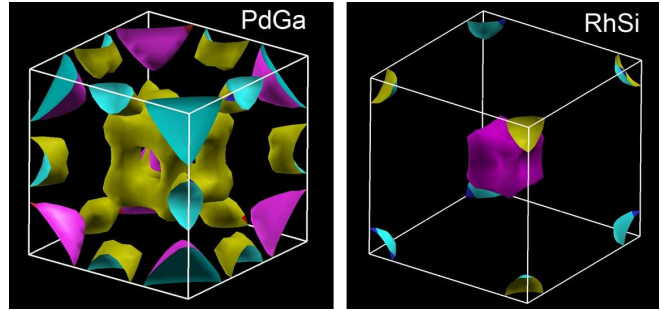}
\caption{Bulk Fermi surfaces of PdGa (left) and RhSi (right).}
\label{FS}
\end{figure}

\textbf{Comparison of the Fermi surfaces of PdGa and RhSi.} The bulk
Fermi surfaces of PdGa and RhSi were calculated using WIEN2k's
\cite{WIEN2k} full-potential linearized augmented plane wave methods
with the Perdew-Burke-Ernzerhof exchange-correlation functional on a
$32 \times 32 \times 32$ $k$-mesh, with account for spin-orbit
coupling and with the lattice parameters taken from
Ref.~\cite{Sessi2020}. The results of these calculations are shown
in Fig.~\ref{FS}.

\clearpage

\end{document}